\documentclass[aps,english,prd,showpacs,showkeys,twocolumn,longbibliography, nofootinbib]{revtex4-1}

\usepackage[english]{babel}
\usepackage[utf8]{inputenc}
\usepackage[T1]{fontenc}
\usepackage{fnpct} 
\usepackage{ulem} 
\usepackage{scrextend} 
\usepackage{enumitem}

\usepackage{amsmath}
\usepackage{amsfonts}
\usepackage{amssymb}
\usepackage{mathtools, stmaryrd}
\usepackage{empheq}
\usepackage{tensor}

\usepackage{fancyvrb}
\usepackage{graphicx}
\usepackage{floatrow}
\usepackage[usenames,dvipsnames]{xcolor}
\usepackage{tabularx}

\usepackage{natbib}
\input{aastexv6.sty} 

\usepackage{url}
\usepackage[pdftex,colorlinks=true, pdfstartview=FitV, linkcolor= Mlink, citecolor=Mlink, urlcolor= Mlink, hyperindex=true, hyperfigures=true]{hyperref}
\hypersetup{linktoc=page}

\input{Short_cut.sty}

\begin{document}

\title{The 1+3-Newton-Cartan system and Newton-Cartan cosmology}
\author{Quentin Vigneron}
\email{quentin.vigneron@ens-lyon.fr}
\email{quvigneron@gmail.com}
\affiliation{Univ Lyon, Ens de Lyon, Univ Lyon1, CNRS, Centre de Recherche Astrophysique de Lyon UMR5574, F–69007, Lyon, France}

\date{\today}

\begin{abstract}
We perform a covariant 1+3 split of the Newton-Cartan equations. The resulting 3-dimensional system of equations, called \textit{the 1+3-Newton-Cartan equations}, is structurally equivalent to the 1+3-Einstein equations. In particular it features the momentum constraint, and a choice of adapted coordinates corresponds to a choice of shift vector. We show that these equations reduce to the classical Newton equations without the need for special Galilean coordinates. The solutions to the 1+3-Newton-Cartan equations are equivalent to the solutions of the classical Newton equations if space is assumed to be
compact or if fall-off conditions at infinity are assumed. We then show that space expansion arises as a fundamental field in Newton-Cartan theory, and not by construction as in the classical formulation of Newtonian cosmology. We recover the Buchert-Ehlers theorem for the general expansion law in Newtonian cosmology.

\end{abstract}


\maketitle

\section{Introduction}
\label{sec::Intro}

The Newton-Cartan theory is a formulation of Newton's theory of gravitation in a 4-dimensional Galilei manifold. The structure on this manifold called a Galilei structure, follows from Newton's ideas of absolute time and absolute space. Similarly to general relativity, the goal behind the Newton-Cartan formulation is to describe the gravitational force with a spacetime connection. The physical equations constraining this connection, called the Newton-Cartan equations, are equations relating the Riemann tensor associated to this connection and the energy content.

The Newton-Cartan theory, originally introduced as a spacetime geometrisation of Newton's theory \cite{1923_Cartan, 1924_Cartan}, has then been developed to study the Newtonian limit (e.g. Ref.~\cite{1976_Kunzle}) and to define post-Newtonian approximations to general relativity (e.g. Refs.~\cite{1997_Dautcourt, 2011_Tichy_et_al}). Ehlers \cite{2019_Ehlers} also proposed a unification of Newton-Cartan theory and general relativity, within his frame theory (see the overview of literature related to the frame theory in Ref~\cite{2019_Buchert_et_al}).

The Newton-Cartan equations can be written as 3-dimensional equations if adapted coordinates are chosen. However, to recover the classical Newton equations, special \textit{Galilean coordinates} are often assumed (e.g. \cite{1990_Dautcourt_a, 1990_Dautcourt_b, 2019_Ehlers}), and an additional constraint on the Riemann tensor, called the ``law of existence of absolute rotation'' by Ehlers \cite{2019_Ehlers} needs to be added so that the Coriolis field does not depend on space.

Newton-Cartan theory was also applied to cosmology in Ref.~\cite{1997_Ruede_et_al} where the authors introduced a fluid model with a 4-velocity similarly to general relativity. They projected the Newton-Cartan equations with respect to this velocity and then assumed homogeneity to derive the expansion law in Galilei coordinates.

The aim of this paper is to present the covariant 1+3-split of the Newton-Cartan equations with respect to the fluid 4-velocity, similarly to the 1+3 and 3+1 split in general relativity. The resulting system of equations will be called \textit{the 1+3-Newton-Cartan equations}. We will show that there is no need to choose specific coordinates, like Galilean coordinates, to recover the classical Newton equations. We will also complete the work of \cite{1997_Ruede_et_al} as we will study space expansion in Newton-Cartan without assuming spatial homogeneity or perturbations.

In Sec.~\ref{sec::NC_Galilei_space} we recall the definition and properties of Galilei spacetimes. Section~\ref{sec::NC_1+3} presents the construction of the 1+3-Newton-Cartan equations. We solve the system of equations in Sec.~\ref{sec::NC_exp} and show that space expansion arises as a fundamental field of the theory. After defining the gravitational field and observers in Sec.~\ref{sec::NC_Observers}, we finally compare the solutions of the 1+3-Newton-Cartan equations to the solutions of the classical Newton equations with a homogeneous deformation in Sec.~\ref{sec::NC_compare}.

The main application we envision for the 1+3-Newton-Cartan equations is to define a \textit{non-Euclidean Newtonian theory}, i.e. a theory which is locally equivalent to Newton's theory of gravitation but in a compact 3-manifold with a non-Euclidean Thurston geometry. This would be a powerful tool to study the effects of a non-Euclidean global geometry on the structure formation in cosmology, as well as on the `backreaction' of these structures on the global expansion of the Universe (see Ref.~\cite{2008_Buchert}): a candidate for the dark energy. We discuss the possibility of defining such a non-Euclidean Newtonian theory with our new framework in Sec.~\ref{sec::NC_disc}.

\section{Galilei spacetimes}
\label{sec::NC_Galilei_space}

\subsection{Notations}
\label{sec::Notation}

We denote a tensor of any type, except scalars, in bold (example: $\T g$). In the case where the type is of importance, a tensor of type $(n,m)$ will feature $n$ over-bars and $m$ under-bars (example: $\Tbb{g}$ for a type $(0,2)$ tensor).

We define the symmetric part $T_{(ab)}$, the antisymmetric part $T_{[ab]}$ and the symmetric traceless part $T_{\langle ab \rangle}$ of a rank-2 tensor $\T T$ as
\begin{align*}
	&T_{(ab)} := \frac{1}{2}\left(T_{ab} + T_{ba}\right) \quad ; \quad T_{[ab]} := \frac{1}{2}\left(T_{ab} - T_{ba}\right) ; \\
	 &T_{\langle ab \rangle} := T_{(ab)} - \frac{T}{D}g_{ab},
\end{align*}
where $\T g$ is the metric of the manifold on which $\T T$ is defined and $D$ the dimension of this manifold.

An antisymmetrisation over three indices is defined as
\begin{equation}
	T_{[abc]} := \frac{1}{3}\left(T_{a[bc]} + T_{c[ab]} + T_{b[ca]}\right).
\end{equation}
An index which should not be included in a antisymmetrisation (over two or three indices) or a symmetrisation is denoted between vertical bars. For instance, in the case of two indices antisymmetrisation:
\begin{equation}
	T_{[a|b|c]} := \frac{1}{2}\left(T_{abc} - T_{cba}\right).
\end{equation}

The Lie derivative on a manifold $\CM$ of a tensor $\T T$ along a vector field $\Tt A$ is denoted $\Lie{\T A} \T T$. The Lie derivative does not commute with the metric, so for instance, for a rank-1 tensor $\T B$,  $\T g(\Lie{\T A} \Tt B, \cdot) \not = \Lie{\T A} \Tb B$. We will then use $\Lie{\T A} B^a$, respectively $\Lie{\T A} B_a$, to denote the coordinate components of $\Lie{\T A} \Tt B$, respectively $\Lie{\T A} \Tb B$.

Then for a vector $\T A$ and a tensor $\T T$ on a manifold $\CM$, we have
\begin{align}
	\tensor[]{{\Lie{\T A}}}{} &{T^{a_1 ...}}_{b_1...} := A^c\nabla_c \tensor{T}{^{a_1}^{...}_{b_1}_{...}} \label{eq::Lie_def} \\
		& + \sum_i {T^{a_1 ...}}_{... \underset{\underset{i}{\uparrow}}{c} ...} \nabla_{b_i} A^c - \sum_j {T^{... \overset{\overset{j}{\downarrow}}{c} ...}}_{b_1 ...} \nabla_{c} A^{a_j}, \nonumber
\end{align}
where $\T \nabla$ is the Levi-Civita connection of $\CM$. The $\tiny \smash{\overset{c}{\underset{i}{\uparrow}}}$ notation means that $c$ is the $i^{\rm th}$ index.

Finally, we denote indices running from 0 to 3 by Greek letters ($\alpha$, $\beta$, $\gamma$, ...) and indices running from 1 to 3 by Roman letters ($a$, $b$, $c$, ...).

\subsection{Galilei structure}
\label{sec::New_from_NC}

This section is largely inspired on the presentation of Galilei spacetimes by K\"unzle (1972)~\cite{1972_Kunzle}.

A \textit{Galilei spacetime} is a 4-dimensional differentiable manifold $\CM$ with a \textit{Galilei structure} $(\Tb\tau, \Ttt h, \Tb\nabla)$, where $\Tb\tau$ is an exact 1-form, $\Ttt h$ is a symmetric (2,0)-tensor of rank 3, with $h^{\alpha\mu}\tau_\mu = 0$, and $\Tb\nabla$ is a connection compatible with $\Tb\tau$ and $\Ttt h$, called a \textit{Galilei connection}:
\begin{equation}
	\nabla_\alpha \tau_\beta = 0 \quad ; \quad \nabla_\gamma h^{\alpha\beta} = 0. \label{eq::NC_def_structure}
\end{equation}
A vector $\Tt u$ is called $\textit{timelike}$ if $u^\mu\tau_\mu = 1$, and an (n,0)-tensor $\T T$ is called \textit{spatial} if $\tau_\mu {T^{... \overset{\overset{\alpha}{\downarrow}}{\mu} ...}} = 0$ for all $\alpha \in \llbracket1,n\rrbracket$.

No spacetime metric, i.e. a symmetric (0,2)-tensor of rank 4, is part of the Galilei structure. Furthermore it is not possible to define a spacetime metric compatible with the connection~\eqref{eq::NC_def_structure} (see Chapter 12 of Ref.~\cite{1973_MTW}). Thus raising and lowering indices is not possible \textit{a priori}. Then when defining new tensors for the first time, we will use the over and under bars notation introduced in Sec.~\ref{sec::Notation} to avoid confusion. Once they have been defined, we will however return to the simpler bold notation.

The exact 1-form $\T \tau$ defines a foliation $\folGR$ in $\CM$, where $\Sigma_t$ are spatial hypersurfaces in $\CM$ defined as the level surfaces of the scalar field $t$, with $\T \tau = \T\dd t$.

From the knowledge of $\T\tau$ and $\T h$, the connection $\T\nabla$ is not unique. Its coefficients ${\Gamma}_{\alpha\beta}^\gamma$ are defined up to a timelike vector $\Tt B$ and a two form $\Tbb\kappa$ as follows:
\begin{equation}
	{\Gamma}_{\alpha\beta}^\gamma = \GB_{\alpha\beta}^\gamma + 2\tau_{(\alpha}\kappa_{\beta)\mu} h^{\mu\gamma}, \label{eq::NC_connection}
\end{equation} 
where
\begin{equation}
	\GB_{\alpha\beta}^\gamma := h^{\gamma\mu}\left(\partial_{(\alpha} \bb{B}_{\beta)\mu} - \frac{1}{2} \partial_\mu \bbB_{\alpha\beta}\right) + B^\gamma \partial_{(\alpha}\tau_{\beta)}, \label{eq::NC_Gamma_B}
\end{equation}
and where $\tensor[^{\T B}]{\Tbb b}{}$ is the projector orthonormal to the vector $\T B$ with
\begin{equation}
	\bbB_{\alpha\mu} B^\mu := 0 \quad ; \quad \bbB_{\alpha\mu}h^{\mu\beta} := {\delta_\alpha}^\beta - \tau_\alpha B^\beta.
\end{equation}
We have the following properties
\begin{equation}
	B^\mu \nabB{B}_\mu B^\alpha = 0 \quad ; \quad h^{\mu[\alpha} \nabB{B}_\mu B^{\beta]} = 0. \label{eq::NC_connection_rel_B}
\end{equation} 
where $\tensor[^{\T B}]{\Tb \nabla}{}$ is the connection associated with the coefficients $\GB_{\alpha\beta}^\gamma$.

The connection coefficients~\eqref{eq::NC_connection} naturally define a Riemann tensor ${R^{\sigma}}_{\alpha\beta\gamma}$ with the formula
\begin{equation}
	{R^{\sigma}}_{\alpha\beta\gamma} := 2 \, \partial_{[\beta}\Gamma^\sigma_{\gamma]\alpha} + 2 \,  \Gamma^\sigma_{\mu[\beta} \Gamma^\mu_{\gamma]\alpha}.
\end{equation}
The Ricci and Bianchi identities are still satisfied for this Riemann tensor:
\begin{align}
	{R^{\sigma}}_{[\alpha\beta\gamma]} &= 0, \\
	\nabla_{[\mu}{R^{\sigma}}_{|\alpha|\beta\gamma]} &= 0.
\end{align}
The definitions~\eqref{eq::NC_def_structure} leads to the following additional relations:
\begin{align}
	\tau_\mu{R^{\mu}}_{\alpha\beta\gamma} &= 0, \\
	h^{\mu(\alpha}{R^{\beta)}}_{\mu\gamma\sigma} &= 0.
\end{align}

The Ricci tensor $R_{\alpha\beta}$ is defined as $R_{\alpha\beta} := {R^{\mu}}_{\alpha\mu\beta}$.

\subsection{Coordinates in a Galilei spacetime}

In this section, we define objects which will be used in the construction of the 1+3-Newton-Cartan equations.

\subsubsection{Adapted coordinate systems.}

A coordinate system $\{x^\alpha\}_{\alpha=0,1,2,3}$, associated to the coordinate basis vectors $\{{\Tt \partial}_\alpha\}_{\alpha=0,1,2,3}$, is said to be adapted to the foliation $\folGR$ if the three vectors $\{{\Tt \partial}_a\}_{a=1,2,3}$ are spatial and ${\Tt \partial}_0$ is timelike\footnote{Actually ${\Tt \partial}_0$ only needs \textit{not} to be spatial. But by convention we take it to be timelike.}. Such coordinates are not unique and are determined up to the spatial vector freedom in the definition of the timelike vector ${\T \partial}_0$.

In any adapted coordinate system
\begin{equation}
	\tau_\alpha = \delta^0_\alpha \quad ; \quad T^{{\alpha_1} ... {\alpha_n}} = T^{{a_1} ... {a_n}} \delta^{\alpha_1}_{a_1} ... \delta^{\alpha_n}_{a_n}, \label{eq::NC_adapted}
\end{equation}
where $\T T$ is a spatial tensor.

\subsubsection{Pull-back.}
\label{sec::NC_pull-back}

The relation~\eqref{eq::NC_adapted} shows that any spatial tensor $\T T$ is totally determined by its components $T^{{a_1} ... {a_n}}$ in an adapted coordinate system. We can then consider that $T^{{a_1} ... {a_n}}$ are the components of a tensor living in a 3-dimensional (hereafter 3D) Riemannian manifold $\Sigma$ whose metric contravariant components are $h^{ab}$, thus defining a pull-back $T^{{\alpha_1} ... {\alpha_n}} \rightarrow T^{{a_1} ... {a_n}}$.

As $\Sigma$ is a Riemannian manifold, indices of tensor components on this manifold can be raised and lowered with the metric $\T h$ on $\Sigma$. We then define $T_{{a_1} ... {a_n}}$ as
\begin{equation}
	T_{{a_1} ... {a_n}} := T^{{c_1} ... {c_n}} h_{c_1a_1} \, ... \, h_{c_na_n}
\end{equation}
where $h_{ab}$ is the inverse matrix of $h^{ab}$ and corresponds to the covariant components of the Riemannian metric on $\Sigma$.

\subsubsection{Classes of adapted coordinates.}

Given a timelike vector $\T u$, one can characterise with respect to this vector any adapted system $\{{\Tt \partial}_\alpha\}_{\alpha=0,1,2,3}$ by introducing a vector $\Tt \beta$ as
\begin{equation}
	\T \beta := {\T \partial}_0 - \T u. \label{NC_shift_vector}
\end{equation}
The vector $\T \beta$ is spatial and is called the \textit{shift vector} of the system $\{{\Tt \partial}_\alpha\}_{\alpha=0,1,2,3}$ with respect to $\T u$. The shift vector defines a class of adapted coordinate systems, denoted $\class{\T \beta}{\T u}$. This class is the set of all adapted coordinate systems whose shift vector with respect to $\T u$ is $\T \beta$. The systems inside a class are related by time-independent  spatial changes of coordinates.

\subsubsection{Spatial covariant derivative.}
\label{sec::NC_spatial_cov_deriv}

In an adapted coordinate system, the spatial projection ${\Gamma}_{ab}^\gamma$ of the connection coefficients~\eqref{eq::NC_connection} are
\begin{equation}
	{\Gamma}_{ab}^\gamma = \delta^\gamma_c h^{cd}\left(\partial_{(a} h_{b)d} - \frac{1}{2} \partial_d h_{ab}\right),\label{eq::NC_connection_spatial}
\end{equation}
for any timelike vector $\T B$ chosen in the relation~\eqref{eq::NC_connection}. This comes from the fact that $\bb{B}_{ab} = h_{ab}$ for any timelike vector $\T B$.

The coefficients~\eqref{eq::NC_connection_spatial} correspond to the coefficients of the Levi-Civita connection $\T D$ of the metric $\T h$ on $\Sigma$. Then the pull-back of $h^{\beta\gamma}\nabla_\gamma T^{{\alpha_1} ... {\alpha_n}}$ on $\Sigma$ with $\T T$ a spatial tensor gives
\begin{equation}
	h^{\beta\gamma}\nabla_\gamma T^{{\alpha_1} ... {\alpha_n}} \rightarrow D^b T^{a_1 ... a_n},
\end{equation}
and the pull-back of the divergence $\nabla_\gamma {T^{\alpha_1 ... \gamma ... \alpha_n}}$ gives
\begin{equation}
	\nabla_\gamma {T^{\alpha_1 ... \gamma ... \alpha_n}} \rightarrow \D_c {T^{a_1 ... c ... a_n}}.
\end{equation}

\section{The 1+3-Newton-Cartan equations}
\label{sec::NC_1+3}

\subsection{The Newton-Cartan equations}

 The Newton-Cartan (NC) equations are:
\begin{align}
	\nabla_\mu T^{\mu\alpha} &=0, \label{eq::NC_Conservation} \\
	R_{\alpha\beta} &= \tau_\alpha \tau_\beta \left( 4\pi G \tau_\mu \tau_\nu T^{\mu\nu} - \Lambda \right), \label{eq::NC_Einstein} \\
	h^{\mu[\alpha}{R^{\beta]}}_{(\gamma\sigma)\mu} &= 0, \label{eq::NC_Kunzle}
\end{align}
where $\Lambda$ is the cosmological constant and $\Ttt T$ is symmetric and corresponds to the stress-energy tensor of the matter.

Equation~\eqref{eq::NC_Conservation} is the energy and momentum conservation; Eq.~\eqref{eq::NC_Einstein} is the equivalent to the Einstein equation and links the geometry of $\CM$ to its energy content; Eq.~\eqref{eq::NC_Kunzle} is the Trautman-K\"unzle condition. \\

\remark{In most of the literature, the condition~\eqref{eq::NC_Kunzle} is called the Trautman condition, citing Trautman (1963) \cite{1963_trautman}\footnote{Paper written in French and available on the website of the Biblioth\`eque Nationale de France at the following web page: \url{https://gallica.bnf.fr/ark:/12148/bpt6k4007z/f639.image}.}
. However Trautman originally gave the condition $h^{\mu[\alpha}{R^{\beta]}}_{\gamma\sigma\mu} = 0$, i.e. without the symmetrisation. This original condition is stronger than~\eqref{eq::NC_Kunzle}. In particular it already implies the proportionality $R_{\alpha\beta} \propto \tau_{\alpha}\tau_{\beta}$, i.e. Eq.~\eqref{eq::NC_Einstein}. Based on a count of the remaining degrees of freedom in the Riemann tensor, K\"unzle (1972) \cite{1972_Kunzle} proposed instead the condition~\eqref{eq::NC_Kunzle}. It has the advantage of still implying the irrotationality of the gravitational field and the closeness of $\T\kappa$, i.e. the original reasons for the introduction of the condition by Trautman in 1963, but without the stronger proportionality implication. That is why we propose to call this condition the \textit{Trautman-K\"unzle condition}.}

For the remainder of this paper, we will only consider that the matter is a fluid described by a timelike vector $\Tt u$ and which stress-energy tensor $\T T$ is\footnote{For an electromagnetic fluid in NC theory, see Ref.~\cite{1976_Kunzle}.}
\begin{equation}
	T^{\alpha\beta} := \rho u^\alpha u^\beta + p h^{\alpha\beta} + 2q^{(\alpha}u^{\beta)} + \pi^{\alpha\beta}, \label{eq::T^u}
\end{equation}
where $\rho$ is the mass density, $p$ the pressure, $\Tt q$ the heat flux and $\Ttt \pi$ the anisotropic stress of the fluid. By definition $q^\mu \tau_\mu = 0$, $\pi^{\mu\alpha} \tau_\mu = 0$ and $\bb{u}_{\mu\nu} \pi^{\mu\nu} = 0$.

\subsection{1+3 split in Newton-Cartan}
\label{sec::1+3-NC}

The basics behind the 1+3 split of the NC equations is to decompose the Ricci tensor along and normal to the fluid velocity $\T u$. To do so, we first introduce the kinematical variables of the fluid in Sec.~\ref{sec::NC_kinematical} and then realise the split 
in Sec.~\ref{sec::NC_1+3-NC}.

This section also completes Sec.~4 in Ref.~\cite{1997_Ruede_et_al} by taking a general fluid, and using all the NC equations.

\subsubsection{The kinematical variables}
\label{sec::NC_kinematical}

Similar to general relativity (GR), we introduce the expansion tensor  $\Ttt \Theta$ and the vorticity tensor $\Ttt\Omega$ of the fluid as the projection orthogonal to the fluid of the 4-velocity gradient $\T\nabla\T u$:
\begin{equation}
	\Theta^{\alpha\beta} :=  {\Pbu}_\nu^{(\beta} h^{\alpha)\mu} \nabla_\mu u^\nu \quad ; \quad \Omega^{\alpha\beta} := {\Pbu}_\nu^{[\beta} h^{\alpha]\mu} \nabla_\mu u^\nu,
\end{equation}
with ${\Pbu}^\beta_\alpha := \bb{u}_{\mu\alpha} h^{\beta\mu} = {\delta_\alpha}^\beta - \tau_\alpha u^\beta$. We denote $\theta := \bbu_{\mu\nu} \Theta^{\mu\nu}$.

We also introduce the acceleration $\TT{\Tt a}{u}$ of the fluid 4-velocity $\T u$ as
\begin{equation}
	\Accu^\alpha := u^\mu\nabla_\mu u^\alpha.
\end{equation}
The tensors $\TT{\T a}{u}$, $\T\Theta$ and $\T\Omega$ are all spatial.

We have the following additional relation
\begin{align}
	\Theta^{\alpha\beta} = -\frac{1}{2}\Lie{\T u} h^{\alpha\beta}. \label{eq::NC_Theta_h}
\end{align}

\remark{Relation~\eqref{eq::NC_Theta_h} was originally introduced by Ref.~\cite{1957_Toupin} as the definition for $\T \Theta$. Note that the relation ``$\bbu_{\alpha\gamma}\bbu_{\beta\sigma}\Theta^{\gamma\sigma} = \frac{1}{2}\Lie{\T u} \bbu_{\alpha\beta}$'' given in Ref.~\cite{1976_Kunzle} is incorrect.}

\subsubsection{1+3-Newton-Cartan equations}
\label{sec::NC_1+3-NC}

In this section we project the NC equations~\eqref{eq::NC_Conservation}-\eqref{eq::NC_Kunzle} with respect to $\T\tau$, $\T u$, $\TT{\T b}{u}$ and $\T h$.

The conservation equation~\eqref{eq::NC_Conservation} projected along $\TT{\T b}{u}$ and $\T\tau$ gives
\begin{align}
		\Lie{\T u}\rho &= - \rho \theta - \nabla_\mu q^\mu, \label{eq::NC_conser_1} \\
	\rho \Accu^\alpha	&= -h^{\mu\alpha}\nabla_\mu P - \nabla_\mu \pi^{\mu\alpha} \label{eq::NC_conser_2} \\
					& \quad \, - \left[\Lie{\T u} q^\alpha + q^\alpha \theta + 2 \bbu_{\mu\nu} q^\mu \left(\Theta^{\nu\alpha} + \Omega^{\nu\alpha}\right)\right]. \nonumber
\end{align}

The NC equation~\eqref{eq::NC_Einstein} projected respectively twice along $\T u$, along $\T u$ and $\T h$, and twice along $\T h$ gives:
\begin{align}
	&\Lie{\T u} \theta = -4\pi G \rho + \Lambda + \nabla_\mu \Accu^\mu \\
	&\quad\quad - \bbu_{\alpha\mu}\bbu_{\beta\nu} \Theta^{\alpha\beta} \Theta^{\mu\nu} +\bbu_{\alpha\mu}\bbu_{\beta\nu} \Omega^{\alpha\beta} \Omega^{\mu\nu}, \nonumber \\
	&h^{\mu\alpha} \nabla_\mu \theta - \nabla_\mu\left(\Theta^{\alpha\mu} + \Omega^{\alpha\mu}\right) = 0, \\
	&h^{\mu\alpha}h^{\nu\beta} R_{\mu\nu} = 0.
\end{align}

In the Trautman-K\"unzle condition~\eqref{eq::NC_Kunzle}, the indices $\alpha$ and $\beta$ are spatial. Then only the indices $\gamma$ and $\sigma$ need to be split along $\T u$ and $\T h$. The condition projected respectively twice along $\T u$, along $\T u$ and $\T h$, and twice along $\T h$ gives:
\begin{align}
	\Lie{\T u} \Omega^{\alpha\beta} &= 4 \bbu_{\mu\nu} \Theta^{\mu[\alpha}\Omega^{\beta]\nu} + h^{\mu[\alpha}\nabla_\mu \Accu^{\beta]}, \\
	h^{\mu[\alpha}\nabla_\mu \Omega^{\beta\gamma]} &= 0, \\
	h^{\mu\gamma}h^{\nu\sigma} h^{\zeta[\alpha}{R^{\beta]}}_{(\mu\nu)\zeta} &= 0. \label{eq::NC_Kunzle_3}
\end{align}

For the system to be closed, the relation~\eqref{eq::NC_Theta_h} needs to be added.

\subsection{The equations}
\label{sec::NC_Pull-back}

The 1+3-NC equations~\eqref{eq::NC_Theta_h}-\eqref{eq::NC_Kunzle_3} are all scalar or spatial equations on $\CM$. By pulling them back they become 3D-equations living on a 3D-Riemannian manifold $\Sigma$. A pull-back is defined for each $\Sigma_t$, i.e. each time $t$. This implies that the geometrical properties (Riemann tensor and metric) of the manifold $\Sigma$, as long as all the other tensors defined on it, are parametrised by the time.

In Sec~\ref{sec::NC_spatial_cov_deriv} we detailed the pull-back of the spatial derivative, but to fully write the 1+3-NC equations as 3D-equations, there remains to pull-back the operator $\Lie{\T u}$ present in the evolution equations. This is done by introducing a class $\class{\T \beta}{\T u}$ of coordinates. Then $\Lie{\T u}$ applied on a spatial tensor $\T T$ becomes, under the pull-back,
\begin{equation}
	\Lie{\T u} T^{\alpha_1 ... \alpha_n} \rightarrow \left(\partial_t - \Lie{\T \beta}\right) T^{a_1 ... a_n},
\end{equation}
where the Lie derivative $\Lie{\T \beta} T^{a_1 ... a_n}$ applied on the spatial components of a spatial tensor corresponds to the Lie derivative on $\Sigma$. \\

The 1+3-Newton-Cartan equations~\eqref{eq::NC_Theta_h}-\eqref{eq::NC_Kunzle_3} on $\Sigma$ can then be written as: evolution equations

\begin{empheq}[box=\fbox]{align}
	\left(\partial_t - \Lie{\T \beta}\right)\rho	&= - \rho \theta - \D_c q^c, \label{eq::NC_conser_1b} \\
	\left(\partial_t - \Lie{\T \beta}\right)h^{ab}	&= -2 \Theta^{ab}, \label{eq::NC_Theta_hb} \\
	\left(\partial_t - \Lie{\T \beta}\right) \theta	&= -4\pi G \rho + \Lambda + \D_c \Accu^c \label{eq::NC_Einstein_ab} \\
									& \quad \, - \Theta^{cd} \Theta_{cd} + \Omega^{cd} \Omega^{cd}, \nonumber \\
	\left(\partial_t - \Lie{\T \beta}\right) \Omega_{ab}	&= \D_{[a} \Accu_{b]}, \label{eq::NC_vorticity_b}
\end{empheq}
and constraint equations
\begin{empheq}[box=\fbox]{align}
	\D^a \theta &= \D_c \left(\Theta^{ac} + \Omega^{ac}\right), \label{eq::NC_Mom} \\
	\D_{[a} \Omega_{bc]}	&= 0, \label{eq::NC_Gauss_Omega} \\
	\tensor[^3]{R}{^{ab}}		&= 0, \label{eq::NC_Einstein_3b} \\
	\tensor[^3]{R}{^{d[abc]}}	&= 0, \label{eq::NC_Kunzle_3b}
\end{empheq}
with
\begin{empheq}[box=\fbox]{flalign}
	\rho \Accu^a			&= -\D^a P -\D_c \pi^{ca} \label{eq::NC_conser_2b} \\
						& \quad \, - \left[\left(\partial_t - \Lie{\T \beta}\right) q^a + q^a \theta + 2 q_c \left(\Theta^{ca} + \Omega^{ca}\right)\right]. \nonumber
\end{empheq}

Equation~\eqref{eq::NC_Kunzle_3b}, coming from Eq.~\eqref{eq::NC_Kunzle_3}, is the Ricci identity for the spatial Riemann tensor, and thus is not a constraint.

In the system~\eqref{eq::NC_Theta_hb}-\eqref{eq::NC_Kunzle_3b}, the shift vector $\T \beta$ is not physical and corresponds to a choice of coordinates.  In the Sec.~\ref{sec::NC_Observers} we will see what choice of $\T \beta$ leads to a Galilean coordinate system. \\

\remark{Ref.~\cite{2020_Vigneron} showed that even in the classical formulation of Newton's theory, the choice of time-parametrised coordinate systems is characterised by a vector, similar to a shift vector.}

\subsection{Discussions}

In the 1+3-NC system, the expansion and vorticity tensors are not defined as the symmetric and antisymmetric gradient of a spatial vector. Instead, they are defined via the constraints~\eqref{eq::NC_Mom} and~\eqref{eq::NC_Gauss_Omega}. This is a major difference with the classical definition of Newton's equations (see Sec.~\ref{sec::NC_class_New}). In Sec.~\ref{sec::NC_Solving} we will see what the consequences for those two tensors are.

Because of these constraint equations the 1+3-NC system is nearly formally equivalent to the 1+3-Einstein system of equations (see Ref.~\cite{2014_Roy}). This was also spotted by the seminal paper on the comparison between Newton's theory and GR \cite{1967_Ellis}. However in that paper, $\T\Theta$ and $\T \Omega$ were not defined via the constraints~\eqref{eq::NC_Mom} and~\eqref{eq::NC_Gauss_Omega}, but directly as the gradients of a spatial vector.

The main difference between the 1+3-NC and 1+3-Einstein systems is the missing of the 1+3-Ricci equation in the former. It is replaced by the flatness of space, i.e. Eq.~\eqref{eq::NC_Einstein_3b}. Ref.~\cite{2020_Vigneron} showed that this results from the fact that this equation becomes a relation for the second order, in $1/c^2$, of the spatial curvature in the Newtonian limit.

The 1+3-NC system does not suffer from the lack of the Hamilton constraint of GR. This is because this constraint is only needed when only the 1+3-Ricci equation is considered, without the Raychaudhuri equation. As the latter equation is present, only the 1+3-Ricci equation is missing. In other words, in GR from the knowledge of the Raychaudhuri equation and the 1+3-Ricci equation, one can derive the Hamilton constraint. \\

\remark{The 1+3-NC system does not feature a dependence on the choice of Galilei connection. This is discussed in Appendix~\ref{app::NC_choice_connection}.}

\section{Space expansion in Newton-Cartan}
\label{sec::NC_exp}

In order to derive the expansion law (in Sec.~\ref{sec::NC_exp_law}), we first need solve the constraint equations~\eqref{eq::NC_Mom} and~\eqref{eq::NC_Gauss_Omega} (in Sec.~\ref{sec::NC_Solving}).

\subsection{Solving the constraint equations}
\label{sec::NC_Solving}

Equation~\eqref{eq::NC_Gauss_Omega} implies that the 2-form $\T \Omega$ is closed, which translates into
\begin{equation}
	\Omega_{ab} = \D_{[a}w_{b]} + \rot_{ab}, \label{eq::NC_sol_omega}
\end{equation}
where $\T\rot$ is a harmonic 2-form on $\Sigma$, i.e. $\D_c\D^c \rot_{ab} = 0$.

The manifold $\Sigma$ being flat with Eq.~\eqref{eq::NC_Einstein_3b}, we can use the decomposition theorem showed by Straumann~\cite{2008_Straumann} to uniquely decompose the expansion tensor into scalar, vector and tensor parts (hereafter SVT decomposition) as
\begin{equation}
	\Theta_{ab} = \chi h_{ab} + \D_{(a}v_{b)} + \Xi_{ab}, \label{eq::NC_decomp}
\end{equation}
with $\chi$ a scalar field and $\T \Xi$ is a transverse-traceless (TT) tensor , i.e. ${\Xi_c}^c := 0$ and $\D_c\Xi^{ca} := 0$. The theorem is valid for a Riemanian metric of constant scalar curvature, with zero traceless Ricci curvature. Fall-off conditions at infinity or compactness of $\Sigma$ also have to be added.

The link between the vectors in the expansion and vorticity tensors is made by the momentum constraint~\eqref{eq::NC_Mom}, which becomes, with the decomposition~\eqref{eq::NC_decomp},
\begin{equation}
	\D_a \chi = \D^c\left(\D_{[c}v_{a]} - \D_{[c}w_{a]}\right),
\end{equation}
using the fact that $\D^c\rot_{ca} = 0$ for a harmonic 2-form.

The right-hand side (hereafter rhs) is divergence free, whereas the left-hand side (hereafter lhs) is vorticity free. Then the Hodge decomposition implies that
\begin{equation}
	\D_a \chi = 0 \quad ; \quad \D^c\left(\D_{[c}v_{a]} - \D_{[c}w_{a]}\right) = 0.
\end{equation}
The first equation implies that $\chi$ is only a function of time. The second equation implies that the 2-form $D_{[c}v_{a]} - D_{[c}w_{a]}$ is co-closed, but as it is also exact, we have $D_{[c}v_{a]} - D_{[c}w_{a]} = 0$. We finally have
\begin{equation}
	\Theta_{ab} = \chi (t) h_{ab} + \D_{(a}v_{b)} + \Xi_{ab} \ ; \ \Omega_{ab} = D_{[c}v_{a]} + \rot_{ab}. \label{eq::NC_Theta_Omega_sol}
\end{equation}

We see that in general, the tensor $\T \Theta$ is not the gradient of a vector, but also features non-zero scalar and tensor parts. The same applies for the vorticity tensor which features a non-zero harmonic part $\T\rot$. The physical role of these terms will be discussed in the next section. \\

\remark{From the NC equations~\eqref{eq::NC_Conservation}-\eqref{eq::NC_Kunzle}, there are no more constraints on the harmonic 2-form $\T \omega$. However, if one derives these equations from a limit of general relativity, an additional constraint appears on the Galilei connection (see Eq.~\eqref{eq::NC_constraint_on_Galilei_struc} in Appendix~\ref{app::NC_Coriolis_zero}), which eventually implies $\T\omega = 0$. Then, \textit{the only physical choice on $\T\omega$ that is compatible with general relativity is $\T\omega = 0$.} This is derived in Appendix~\ref{app::NC_Coriolis_zero}. As we did not consider the NC theory as a limit of general relativity in this paper, we will keep $\T\omega \not= 0$.}

\subsection{Space expansion in the Newton-Cartan theory}
\label{sec::NC_exp_law}

The expansion rate $H_\Sigma(t)$ of $\Sigma$ is defined as
\begin{equation}
	H_\Sigma(t)	:= \frac{1}{3} \frac{\partial_t \mathcal V_\Sigma}{\mathcal V_\Sigma} = \frac{1}{3}\Saverage{\theta}(t), \label{eq::NC_exp}
\end{equation}
where $\Saverage{\bullet} := \frac{1}{V_\Sigma}\int_\Sigma \bullet \sqrt{\mathrm{det}(h_{ab})}\dd^3 x$ is the spatial average over the whole manifold $\Sigma$ and $V_\Sigma := \int_\Sigma \sqrt{\mathrm{det}(h_{ab})}\dd^3 x$ is the volume of $\Sigma$.

The definition~\eqref{eq::NC_exp} of the expansion rate shows that
\begin{equation}
	H_\Sigma(t) = \chi(t) \label{eq::NC_H=chi}
\end{equation}
and thus that \textit{$\chi$ is the space expansion rate of $\Sigma$.}

As we explained in the previous section, the scalar $\chi$ enters in the SVT decomposition of the expansion tensor $\T\Theta$. This tensor, along with the vorticity tensor, characterises the fluid. Then any part of the decomposition of $\T\Theta$ corresponds to a physical fundamental field characterising the fluid. In particular, Eq.~\eqref{eq::NC_H=chi} shows that \textit{the expansion}, through $\chi$, \textit{is a fundamental physical field} of the 1+3-NC system~\eqref{eq::NC_conser_1b}-\eqref{eq::NC_conser_2b}.

This system does not explicitly feature an evolution equation for the scalar $\chi$. Such an equation can be obtained by taking the spatial average of the Raychaudhuri equation~\eqref{eq::NC_Einstein_ab}, which gives
\begin{align}
	3\left[\left(\partial_t - \Lie{\T \beta}\right)  H_\Sigma + H_\Sigma^2\right] = &- 4\pi G \Saverage{\rho} + \Lambda \label{eq::NC_fried_gen} \\
	&- \Saverage{\Xi_{cd}\Xi^{cd}} + \Saverage{\rot_{cd}\rot^{cd}}. \nonumber
\end{align}
To get this equation we used the fact that divergences averaged over a compact domain are zero due to Stokes' theorem.
Equation~\eqref{eq::NC_fried_gen} is the Friedmann equation for the acceleration rate of $H_\Sigma$ with two additional terms being $\Saverage{\Xi_{cd}\Xi^{cd}}$ and $\Saverage{\rot_{cd}\rot^{cd}}$.

While we have an evolution equation for $\chi$, this is not the case for the TT tensor $\T\Xi$ and the harmonic 2-form $\T\rot$ which are totally free, in their spatial and time dependence. We call $\T\Xi$ the \textit{global shear}, and $\T\rot$ the \textit{Coriolis field}. The reason for this name will be given in Sec.~\ref{sec::NC_test_obs}.

\section{Gravitational field and observers in Newton-Cartan theory}
\label{sec::NC_Observers}

\subsection{The gravitational field}
\label{sec::NC_gal_def}

In the literature concerning the Newton-Cartan theory, the gravitational field is often defined using the coefficients $\tensor{\Gamma}{^\gamma_\alpha_\beta}$ of the Galilei connection as (e.g. Refs~\cite{2019_Ehlers,1990_Dautcourt_a})
\begin{equation}
	g^a := \tensor{\Gamma}{^a_0_0}. \label{eq::NC_grav_Gamma}
\end{equation}
This definition is however valid only in a specific adapted coordinate system (defined by a time vector $\tensor[^{\T g}]{\T{\partial_t}}{}$) such that
\begin{equation}
	\tensor[^{\T g}]{\T{\partial_t}}{} := \T B, \label{eq::NC_partial_t_B_g}
\end{equation} 
where $\T B$ is the vector freedom in the definition of the Galilei connection of Eq.~\eqref{eq::NC_connection}. The definition~\eqref{eq::NC_grav_Gamma} implies that the gravitational field is the opposite of the 4-acceleration of $\T B$, with $B^\mu\nabla_\mu B^\alpha = -g^\alpha$, and that it is solution of the cosmological Poisson equation \{Eq.~(16) in Ref.~\cite{1990_Dautcourt_a}\}.

In this paper we are interested in giving a purely coordinate independent definition of $\T g$. We propose the following definition:
\begin{equation}
	g^a := \left(\partial_t - \Lie{\T \beta}\right) v^a + 2v^c\left({\Theta_c}^a + {\Omega_c}^a\right) - v^c\D_c v^a - \Accu^a. \label{eq::NC_def_g}
\end{equation}
In the following and in Sec.~\ref{sec::NC_gal_obs} we will see that it is coherent with the standard definition~\eqref{eq::NC_grav_Gamma}. \\

Our definition~\eqref{eq::NC_def_g} implies that the vorticity equation~\eqref{eq::NC_vorticity_b} becomes
\begin{equation}
	\D_{[a} g_{b]} = -\left(\partial_t - \Lie{\T \beta + \T v}\right)\rot_{ab}, \label{eq::NC_vort_g}
\end{equation}
and the Raychaudhuri equation~\eqref{eq::NC_Einstein_ab} becomes
\begin{align}
	\D_c g^c	= &-4\pi G \rho + \Lambda - 3\left[ \left(\partial_t - \Lie{\T \beta}\right) \chi + \chi^2\right] \label{eq::NC_Poisson_g_homo} \\
			&- \Xi_{cd}\Xi^{cd} + \rot_{cd}\rot^{cd}, \nonumber
\end{align}
which can be rewritten, using the expansion law~\eqref{eq::NC_fried_gen}, as follows
\begin{align}
	\D_c g^c	= &-4\pi G \widehat\rho - \widehat{\Xi_{cd}\Xi^{cd}} + \widehat{\rot_{cd}\rot^{cd}}, \label{eq::NC_Poisson_av}
\end{align}
where $\widehat{f} := f - \Saverage{f}$ with $f$ a scalar field. 

Equations~\eqref{eq::NC_vort_g} and~\eqref{eq::NC_Poisson_g_homo} are equivalent to Eqs.~(12) and (16) in Ref.~\cite{1990_Dautcourt_a} implying that the vector field $\T g$ we defined in Eq.~\eqref{eq::NC_def_g}  can indeed be interpreted as the gravitational field. This is further confirmed in Sec~\ref{sec::NC_gal_obs}.

The definition~\eqref{eq::NC_def_g} could be used as a covariant definition of the gravitational field in GR, as it uses the same kinematical variables $\T\Theta$ and $\T\Omega$. However it requires the knowledge of the vector $\T v$ defined from $\T\Theta$ using the SVT decomposition. This decomposition is \textit{a priori} not possible in general in GR as the curvature orthogonal to the fluid is not necessarily zero. Thus it seems non-trivial to adapt the definition~\eqref{eq::NC_def_g} in this theory. \\

\remark{Equations~\eqref{eq::NC_vort_g} and~\eqref{eq::NC_Poisson_av} were also obtained  from NC by \cite{2019_Ehlers,1990_Dautcourt_b}. But these studies neither assume expansion, nor global shear and therefore do not have the term $\widehat{\Xi_{cd}\Xi^{cd}}$ and the averages given by the operator $\widehat{\bullet}$ in the gravitational field source equation~\eqref{eq::NC_Poisson_av}.}

\subsection{General observers}

An \textit{observer} in the NC theory is described by a timelike vector $\T \Ob$. A choice of coordinates can be associated to a choice of observer whose timelike vector is the time basis vector, i.e. $\T \Ob = \T{\partial_t}$. In this sense, solving the NC equations~\eqref{eq::NC_conser_1b}-\eqref{eq::NC_Kunzle_3b} in such coordinates corresponds to solving the dynamics of the fluid $\T u$ with respect to the observer $\T \Ob$.

For a general observer $\T \Ob \not= \T u$, and we define the spatial vector $\T V$ as
\begin{equation}
	\T V := \T \Ob - \T u + \T v.
\end{equation}
We have $\T V = \T \beta + \T v$. The acceleration $\TT{\T a}{\Ob}$ of this observer can then be written as
\begin{align}
	\Acc{\Ob} &= - g^a + \left(\partial_t - \Lie{\T \beta + \T v}\right) V^a + V^c\D_c V^a \\
	&\quad \, + 2V^c\left({\Theta_c}^a + {\Omega_c}^a - \D_cv^a\right), \nonumber
\end{align}
which simplifies into
\begin{align}
	\Acc{\Ob} = - g^a + \partial_t V^a + V^c\D_c V^a+ 2V^c\left(\chi {\delta_c}^a + {\Xi_c}^a + {\rot_c}^a\right). \label{eq::NC_acc_decomp}
\end{align}

\subsection{Galilean observers}
\label{sec::NC_gal_obs}

When the observer is chosen such that $\T V = 0$, then
\begin{equation}
	\Acc{\Ob} = - g^a.
\end{equation}
This corresponds to an observer whose acceleration is the opposite of the gravitational field created by the fluid $\T u$. Such an observer is called a \textit{Galilean observer}. A class of coordinates associated to a Galilean observer is called a \textit{Galilean class}.

With the above definition, there seems to be a unique Galilean observer, i.e. the observer $\T \Ob = \T u - \T v$. However, this is true only if $\T v$ is unique from the knowledge of $\T \Theta$. This is not the case as $\T v$ is defined up to a spatial vector $ \T A$ whose spatial gradient $\D_a A_b$ is zero. For a flat space, the solution to the equation $\D_a A_b = 0$ is a ``constant'' spatial vector, i.e. corresponding to a global translation. Thus $\T v$ is defined up to a global translation, and therefore a Galilean observer is also defined up to a global translation.

However once we choose the vector $\T v$ this fixes the Galilean observer. We denote this observer with the vector $\T G$, where $\T G := \T u - \T v$. Then the vector $\T V = \T o - \T G$ of a general observer $\T o$ corresponds to its spatial velocity with respect to the Galilean observer $\T G$. Therefore, the vector $\T v = \T u - \T G$ is \textit{the spatial velocity of the fluid with respect to the Galilean observer}.

\begin{figure}[h]
	\centering
	\includegraphics[width=\textwidth]{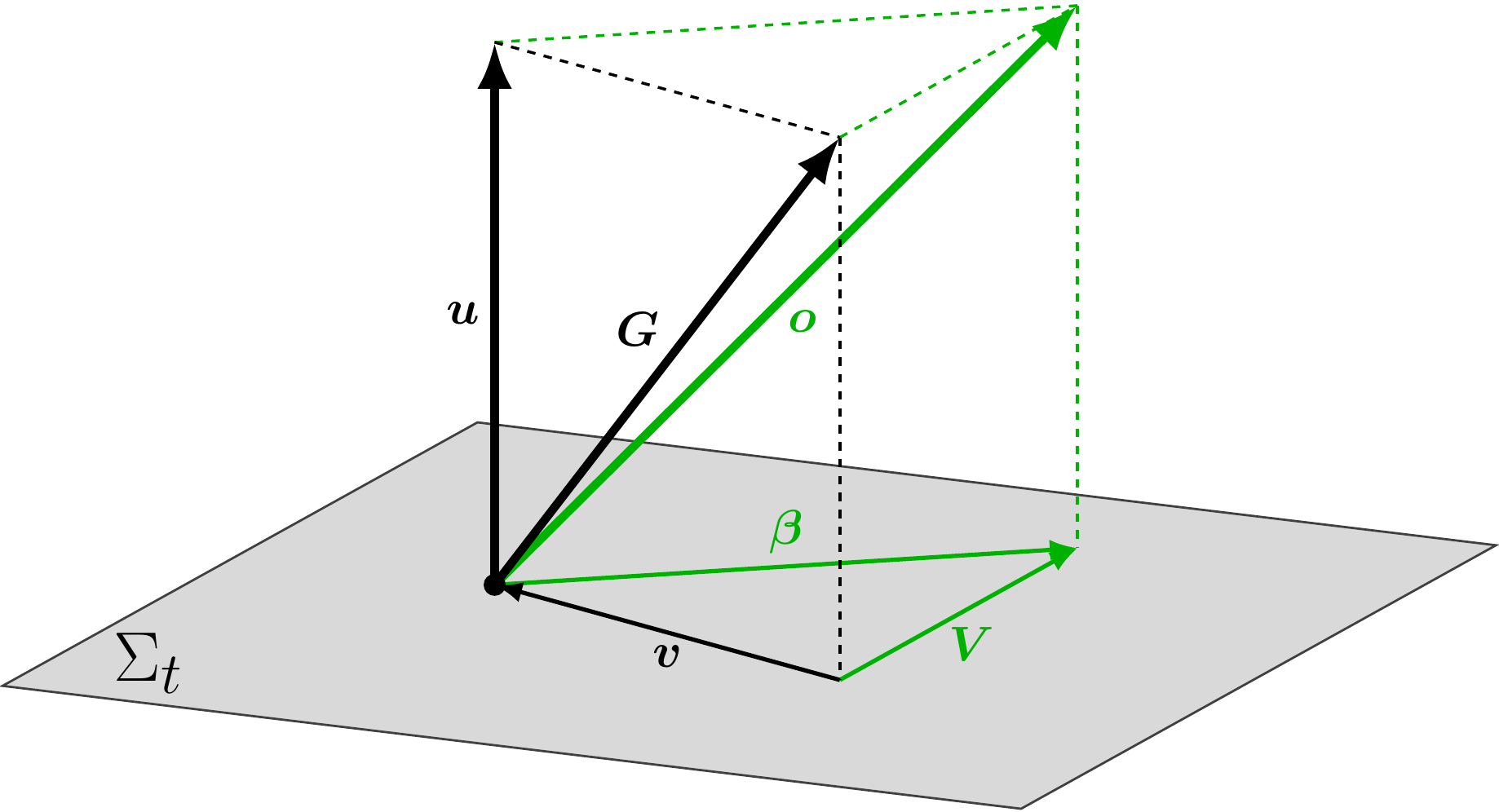}
	\caption{Representation of the different vectors involved in the definition of the fluid $\T u$, the Galilean observer $\T G$ and a general observer $\T o$. These vectors are represented with respect to a slice $\Sigma_t$ of the foliation $\folGR$. In black are the vectors defining the fluid; in green are the vectors defining a general observer. Note that because there is no global metric in the structure of the embedding Galilei manifold $\CM$, the orthogonality on the figure between the timelike vector $\T u$ and the slice $\Sigma_t$ has no signification and is just a representation convention.}
	\label{fig::NC_fig1}
\end{figure}

General and Galilean observers are represented with respect to the fluid in Fig.~\ref{fig::NC_fig1}. \\

\remark{If the Galilei connection is chosen such that $\T B := \T G$, then in the coordinate class $\class{-\T v}{\T u}$, i.e. $\T{\partial_t} := \T G$, we retrieve the usual definition~\eqref{eq::NC_grav_Gamma} of Refs.~\cite{2019_Ehlers,1990_Dautcourt_a} for the gravitational field.}

\subsection{Test observers}
\label{sec::NC_test_obs}

We define a test observer with timelike vector $\T T$ as a geodesic observer, i.e. $\TT{\T a}{T} = 0$. The equation of motion of these observers with respect to the Galilean observer, i.e. the evolution equation for $\T V$, is
\begin{equation}
	\partial_t V^a + V^c\D_c V^a = g^a - 2V^c\left( \chi {\delta_c}^a + {\Xi_c}^a + {\rot_c}^a\right). \label{eq::NC_test_part}
\end{equation}
This corresponds to the second law of Newton with the velocity acceleration on the lhs and three non-inertial terms on the rhs: $- 2 \, \chi V^a$, $- 2V^c{\Xi_c}^a$ and $- 2V^c{\rot_c}^a$. The first one corresponds to an expansion force; the second one to an anisotropic force resulting from the global shear; the third one is a Coriolis force created by the Coriolis field $\T\rot$.

The term $- 2V^c{\rot_c}^a$ corresponds to the Coriolis force created by a global rotation only if the Coriolis field components $\rot_{ab}$ are constants in Cartesian coordinates. As $\T\rot$ is harmonic along with compactness or fall-off conditions at infinity, this is the case in the present paper. This confirms the result of Ref.~\cite{1990_Dautcourt_b}. Note that Ref.~\cite{2019_Ehlers} does not suppose closing conditions, but instead adds an additional constraint to the NC system in order for $\T\rot$ to be a global rotation: the ``law of existence of absolute rotation'' \{Eq.~(23) in Ref.~\cite{2019_Ehlers}\}. However, this equation does not have a relativistic equivalent, and therefore cannot be obtained from a Newtonian limit.

\section{Comparison with the classical formulation}
\label{sec::NC_compare}

We recall the classical Newton equations with a homogeneous deformation in Sec.~\ref{sec::NC_class_New}, and compare them with the 1+3-NC equations in Sec.~\ref{sec::NC_NC_vs_class}.

\subsection{Homogeneous deformation in the classical Newton theory}
\label{sec::NC_class_New}

The classical Newton equations, in their kinematical forms, are
\begin{align}
	\left(\partial_t + \Lie{\T v}\right) \rho		&= - \rho \theta, \label{eq::NC_class_1} \\
	\left(\partial_t + \Lie{\T v}\right) \theta	&= -4\pi G \rho + \Lambda + D_ca^c_{\not= \rm grav} \nonumber \\
									& \quad \ - \Theta_{cd}\Theta^{cd} + \Omega_{cd}\Omega^{cd} , \label{eq::NC_class_2} \\
	\left(\partial_t + \Lie{\T v}\right)\omega_{cd}	&= \D_{[a} (a_{\not= \rm grav})_{b]}, \label{eq::NC_class_3}
\end{align}
where $\T{a}_{\not= \rm grav}$ corresponds to the non-gravitational acceleration, and with
\begin{align}
	\Theta_{ab} := D_{(a} \vN_{b)} \quad ; \quad \Omega_{ab} := D_{[a}\vN_{b]}, \label{eq::NC_class_4}
\end{align}
where $\T\D$ is a flat connection not depending on time.

For the above system to be well defined, closing conditions need to be added. There are two possibilities:
\begin{enumerate}[label=(\roman*)]
	\item the system is isolated, and fall-off conditions are taken at infinity, i.e. $\rho \xrightarrow{r \rightarrow \infty} 0$ and $\T v \xrightarrow{r \rightarrow \infty} \T 0$.
	\item the velocity $\T v$ is decomposed into a \textit{homogeneous deformation} velocity $\T{v}_{\text{H}}$ with ${v}_{\text{H}}^a := {H_c}^{a} x^c$ (in Cartesian coordinates) where ${H^c}_{a} $ are functions of $t$ only, and a peculiar-velocity $\T P$: we have $\T v = \T{v}_{\text{H}} + \T P$. The peculiar-velocity is periodically defined on $\mathbb{R}^3$, with the constraint $\int P^a \dd x = 0$ in Cartesian coordinates, where the integral is taken over the volume defined by the periodic conditions.
\end{enumerate}
The first case is used for astrophysical flows, and the second one for cosmological flows. In the second choice, the trace ${H_c}^{c} =: 3H$ corresponds to the volume expansion of the periodic boundary conditions imposed on $\T P$; the symmetric traceless part of ${H_c}^{a}$ corresponds to the anisotropic expansion of these conditions. Once the homogeneous deformation is introduced, the spatial velocity which describes the fluid is considered to be the peculiar-velocity $\T P$.

The vector $\T{v}_{\text{H}}$, and consequently $\T v$, is defined on $\mathbb{R}^3$, implying that the periodic boundary conditions imposed on $\T P$ only effectively define a compact space. Therefore \textit{the homogeneous deformation is only a construction allowing for the description of expansion}, in an effective compact space, in the classical formulation of Newton's theory. In other words, the expansion rate $H$ quantifies the expansion of the boundary conditions imposed by construction on the peculiar-velocity, and not the expansion of a compact space. \\

\remark{Newtonian cosmological simulations use a trace homogeneous deformation, called a \textit{Hubble flow}, to ``simulate'' expansion while using Newton's theory. In this case the velocity $\T v$ is
\begin{equation}
	v^a = H(t)x^a + P^a,
\end{equation}
in Cartesian coordinates.}

\subsection{1+3-NC VS Classical Newton}
\label{sec::NC_NC_vs_class}

The expansion and vorticity tensors in the classical theory with homogeneous deformation can be written as
\begin{equation}
	\Theta_{ab} = H(t) h_{ab} + \D_{(a}P_{b)} + H_{\langle ab \rangle} \ ; \ \Omega_{ab} = D_{[c}P_{a]} + H_{[ ab ]}. \label{eq::NC_Theta_Omega_class}
\end{equation}
These expressions are equivalent to those in NC [Eq.~\eqref{eq::NC_Theta_Omega_sol}], if we make the following associations:
\begin{equation}
\T P \rightarrow \T v  \ ; \quad H \rightarrow \chi \ ; \quad H_{\langle ab \rangle} \rightarrow \Xi_{ab} \ ; \quad H_{[ ab ]} \rightarrow \rot_{ab}. \label{eq::NC_assoc}
\end{equation}
As $H_{ab}$ is only a function of time in Cartesian coordinates, this implies: firstly, that $H_{\langle ab \rangle}$ is divergence-free and falls into the class of TT-tensors, and secondly, that $H_{[ ab ]}$ is harmonic. This justifies the last two associations.

Furthermore, as the Raychaudhuri and the vorticity conservation equations are formally the same in both formulations [Eqs.~\eqref{eq::NC_Einstein_ab} and~\eqref{eq::NC_vorticity_b} in NC; Eqs.~\eqref{eq::NC_class_2} and~\eqref{eq::NC_class_3} in classical Newton], then the equations governing the peculiar-velocity $\T P$ and the homogeneous deformation are exactly the 1+3-NC equations in a compact space with the association~\eqref{eq::NC_assoc}. This shows that the solutions to the classical Newton equations with the addition of a homogeneous deformation are equivalent to the solutions of the 1+3-NC equations.

As an example, the expansion law we derived in NC [Eq.~\eqref{eq::NC_fried_gen}] is the same as the one in the classical theory, given by the Buchert-Ehlers theorem \{Eq.~(B.6) in Ref.~\cite{1997_Buchert_et_al}\}. This theorem which states, in particular, that for a Hubble flow the expansion is given by the Friedmann equation, has been retrieved in the present paper from the NC theory. \\

\remark{From the association $\quad H_{\langle ab \rangle} \rightarrow \Xi_{ab}$ we can interpret $\Xi$ as an anisotropic expansion if its components are spatially constant in Cartesian coordinates. However, as the TT condition on $\T\Xi$ is more general (see Ref.~\cite{2018_Tafel} for the study of the TT tensors in flat spaces) and \textit{a priori} allows for non-constant components in Cartesian coordinates, we expect the physical effects of this tensor to be more than an anisotropic expansion. If this is indeed the case, the NC theory would be slightly more general than the classical Newton theory.}

\section{On non-Euclidean Newtonian theories}
\label{sec::NC_disc}

One of the candidate for the dark energy, i.e. the recent acceleration of the expansion, is the effect, called `backreaction', of inhomogeneities on this expansion (see Ref.~\cite{2008_Buchert}). The Buchert-Ehlers theorem we retrieved shows that, in Newton's theory, an isotropic expansion is necessarily given by the Friedman equation [Eq.~\eqref{eq::NC_fried_gen} with $\T\Xi = 0$ and $\T\omega = 0$], and therefore needs at least a cosmological constant to feature an acceleration.

This theorem is therefore a major result of cosmology as it tells us that if our Universe has an Euclidean geometry and is well described on small scales by Newton's gravitation, dark energy could not be explained by inhomogeneities (see Refs.~\cite{2017_Kaiser, 2018_Buchert}). This result is however limited to Euclidean geometries, and therefore needs to be generalised for non-Euclidean ones. This requires the definition of a non-Euclidean Newtonian theory (NEN), i.e. \textit{a theory locally equivalent to Newton's theory but defined in a non-Euclidean 3-manifold}. In addition to generalising the Buchert-Ehlers theorem, such a theory would be a powerful tool to study the effects of topology on the structure formation.

Several attempts have been made to define a NEN theory \cite{2009_Roukema_et_al, 2020_Barrow}. Both rely on keeping formally the Poisson equation of the classical Newton theory and assuming that the metric is not flat anymore, but has a curvature corresponding to the desired topology. A first issue with this approach is that it is purely formal, and is not justified from GR. Furthermore, it depends on the equations we consider: in Newton, one can derive the Poisson equation from the Raychaudhuri equation and vice versa, but if we assume a non-flat metric, this is not the case anymore. So the gravitational field obtained with a non-flat metric, i.e. in a non-Euclidean geometry, would be different depending on the equation we kept to define the NEN theory.

In this paper, we showed that Newton's theory can be written as a 3D-system which is formally equivalent to the 1+3-Einstein system (apart for the 1+3-Ricci equation), and has zero Ricci curvature. Then if one wants to keep a formal approach in the definition of a NEN theory, and have a possible justification from GR, we think that this should be done with the 1+3-NC system. 

This was indirectly proposed by K\"unzle \cite{1976_Kunzle}\footnote{K\"unzle claims to give references for the modification~\eqref{eq::NC_Einstein_R}, but these are unrelated to this equation. We can therefore consider Ref.~ \cite{1976_Kunzle} to be the first occurrence of this modification.}, who modified the NC equation~\eqref{eq::NC_Einstein} as follows
\begin{equation}
	R_{\alpha\beta} - \frac{\CR(t)}{3}\bbu_{\alpha\beta} = \tau_\alpha \tau_\beta \left( 4\pi G \tau_\mu \tau_\nu T^{\mu\nu} - \Lambda \right), \label{eq::NC_Einstein_R}
\end{equation}
with $\CR(t)$ a spatial constant. This modification implies the same 1+3-NC equations of the present paper, with Eq.~\eqref{eq::NC_Einstein_3b} replaced by
\begin{equation}
	\tensor[^3]{R}{^{ab}}	= \frac{\CR}{3}h_{ab}.
\end{equation}
This corresponds to our proposed formal approach to define a NEN theory. However solving this new system, especially deriving the expansion law, is beyond the scope of this paper.

This approach being only formal, it lacks of clear justifications from GR. These could be found by adapting the Newtonian limit from GR (e.g. \cite{1976_Kunzle}) to allow for non-zero spatial curvature at the limit.

\section{Conclusion}

This paper aimed at presenting the equations resulting from a covariant 1+3-split of the Newton-Cartan equations, called the \textit{1+3-Newton-Cartan equations} [Eqs.~\eqref{eq::NC_conser_1b}-\eqref{eq::NC_conser_2b}], and the solutions to these equations. The main results are:
\begin{enumerate}[label=(\roman*)]
	\item The 1+3-Newton-Cartan equations have the same algebraic structure as the 1+3-Einstein evolution \textit{and} constraint equations (apart the 1+3-Ricci equation). In particular, as in the relativistic theory, a choice of spatial coordinates in Newton-Cartan corresponds to a choice of shift vector.
	\item We give a covariant definition of the gravitational field [Eq.~\eqref{eq::NC_def_g} in Sec.~\ref{sec::NC_gal_def}],
	\item When solving the constraint equations, the space expansion arises as a fundamental physical field in the theory. This contrasts with the classical theory of Newton where the expansion is only a construction, called homogeneous deformation.
	\item The solutions to the 1+3-Newton-Cartan equations are equivalent to the solutions of the classical Newton equations with a homogeneous deformation, assuming fall-off conditions at infinity or spatial compactness.
	\item We retrieve the Buchert-Ehlers theorem (Ref.~\cite{1997_Buchert_et_al}) in the Newton-Cartan theory, with the expansion law~\eqref{eq::NC_fried_gen}.
	\item We show that the Coriolis field should be zero if one derives the Newton-Cartan theory as a limit (e.g. Ref.~\cite{1976_Kunzle}) of general relativity (see Appendix~\ref{app::NC_Coriolis_zero}).
\end{enumerate}

We also discussed the possibility of defining a non-Euclidean Newtonian theory from the 1+3-Newton-Cartan equations.

\section*{Acknowledgements}

This work is part of a project that has received funding from the European Research Council (ERC) under the European Union’s Horizon 2020 research and innovation programme (Grant agreement ERC advanced Grant 740021–ARTHUS, PI: Thomas Buchert). I was supported by a ‘sp\'ecifique Normalien’ PhD Grant from the \'Ecole Normale Sup\'erieure de Lyon. I thank Étienne Jaupart, Pierre Mourier and Léo Brunswic for a lot of valuable discussions on technical parts. I would also like to thank Thomas Buchert and Boudewijn Roukema for discussions and comments on the manuscript, Guillaume Laibe for comments on the final manuscript, and Madeleine Pham-Thanh for corrections on the final manuscript.

\appendix

\section{Choice of Galilei connection}
\label{app::NC_choice_connection}

Throughout this paper we made no assumption on the choice of Galilei connection, i.e. choice of $\T B$ and $\T\kappa$ in Eq.~\eqref{eq::NC_connection}. However once we wrote the 1+3-NC equations and pulled them back on $\Sigma$, the Galilei connection disappears. Then the intrinsic freedom ont the definition of this connection disappears too. As in most of the literature on NC, reasoning is often made with $\T B$ and $\T\kappa$, we detail in the present section the relation between these two tensors and the kinematical variables.

We have
\begin{align}
	\Theta^{\alpha\beta}		&= h^{\mu(\alpha}\nabB{B}_\mu u^{\beta)}, \\
	\Omega^{\alpha\beta}	&= h^{\mu[\alpha}\nabB{B}_\mu u^{\beta]} + h^{\mu\alpha}\kappa_{\mu\nu}h^{\beta\nu}, \label{eq::NC_omega_decomp_connection} \\
	\Accu^\alpha			&= u^\mu\nabB{B}_\mu u^\alpha + 2u^\mu\kappa_{\mu\nu}h^{\alpha\nu}.
\end{align}

If we choose $\T B = \T u$, then using Eq.~\eqref{eq::NC_connection_rel_B} we have
\begin{align}
	\Theta^{\alpha\beta}		&= h^{\mu(\alpha}\nabB{u}_\mu u^{\beta)}, \\
	\Omega^{\alpha\beta}	&= h^{\mu\alpha}\kappa_{\mu\nu}h^{\beta\nu}, \\
	\Accu^\alpha			&= 2u^\mu\kappa_{\mu\nu}h^{\alpha\nu}.
\end{align}
In this choice of connection, we see that the spatial projection of $\T\kappa$ is the vorticity of the fluid. Then this projection cannot be taken to zero as this would be a physical restriction to the fluid.

Only if $\T B \not= \T u$, one is allowed to take $h^{\mu\alpha}\kappa_{\mu\nu}h^{\beta\nu} = 0$ without loss of generality. But in any case, the tensors $\T B$ and $\T \kappa$ do not appear in the 1+3-NC equations on $\Sigma$, and thus their choice, in addition to having no physical implications, is not relevant to the solving of these equations. Only the choice of adapted coordinates via $\T \beta$ defining the partial time derivative, i.e. the choice of observer, plays a role in Eqs.~\eqref{eq::NC_conser_1b}-\eqref{eq::NC_conser_2b}.

\section{$\T\omega = 0$ from general relativity.}
\label{app::NC_Coriolis_zero}

\subsection{The Newtonian limit}

We consider a manifold $\CM$ and a Lorentzian structure ($\T g$, $\overset{\T g}{\nabla}$) on this manifold, where $\T g$ is a Lorentzian metric and $\overset{\T g}{\nabla}$ the Levi-Civita connection associated to $\T g$.

The limit allowing for the recovering of the NC equations from general relativity is based on a Taylor expansion of $\T g$ in powers of $\lambda := 1/c^2$ so that (e.g. Ref.~\cite{1976_Kunzle})
\begin{align}
	g^{\alpha\beta} &= h^{\alpha\beta} + \mathcal{O}(\lambda), \label{eq::NC_g^ab_lim} \\
	g_{\alpha\beta} &= -\frac{1}{\lambda} \tau_\alpha \tau_\beta + \mathcal{O}(1), \label{eq::NC_g_ab_lim} 
\end{align}
where $\T \tau$ is an exact 1-form and $\T h$ is a (2-0)-tensor of rank 3.

From Eqs.~\eqref{eq::NC_g^ab_lim}  and~\eqref{eq::NC_g_ab_lim}, the connection $\overset{\T g}{\nabla}$ can be developed in powers of $\lambda$ (see the first equation on page 452 in Ref.~\cite{1976_Kunzle}), giving
\begin{equation}
	\overset{\T g \ \ \ }{\Gamma^\gamma_{\alpha\beta}} = \GB_{\alpha\beta}^\gamma + \tau_\alpha\tau_\beta h^{\gamma\mu}\partial_\mu \phi + \mathcal{O}(\lambda), \label{eq::NC_gamma_lim}
\end{equation}
where $\phi$ is an arbitrary scalar field and $\GB_{\alpha\beta}^\gamma$ is defined by Eq.~\eqref{eq::NC_Gamma_B} with $\T B$ a vector satisfying $B^\mu\tau_\mu = 1$.

We see that the leading order of the Lorentzian connection corresponds to a Galilei connection. Then from Eqs.~\eqref{eq::NC_g^ab_lim}-\eqref{eq::NC_gamma_lim}, the Lorentzian structure ($\T g$, $\overset{\T g}{\nabla}$) is a Galilei structure $(\T\tau, \T h, \T\nabla)$ at leading order. It is then possible to develop the Einstein equation, the momentum conservation and the Bianchi identity at leading order to obtain the Newton-Cartan system~\eqref{eq::NC_Conservation}-\eqref{eq::NC_Kunzle}.

\subsection{Constraint on $\T\omega$}

Contrary to the Galilei structure considered in this paper, the one obtained from general relativity is constrained: Eq.~\eqref{eq::NC_gamma_lim} implies the contraint
\begin{equation}
	2\tau_{(\alpha}\kappa_{\beta)\mu} h^{\mu\gamma} = \tau_\alpha\tau_\beta h^{\gamma\mu}\partial_\mu \phi. \label{eq::NC_constraint_on_Galilei_struc}
\end{equation}
Then $h^{\mu\alpha}\kappa_{\mu\nu}h^{\beta\nu} = 0$, which from Eq.~\eqref{eq::NC_omega_decomp_connection}, implies
\begin{equation}
	\Omega^{\alpha\beta}	= h^{\mu[\alpha}\nabB{B}_\mu u^{\beta]}.
\end{equation}
Then using Eq.~\eqref{eq::NC_connection_rel_B}, we have
\begin{equation}
	\Omega^{\alpha\beta}	= h^{\mu[\alpha}\nabB{B}_\mu \left(\T u - \T B\right)^{\beta]}.
\end{equation}
As $\T p := \T u - \T B$ is spatial, then $\Omega^{ab} = D^{[a}p^{b]}$ and $\T\Omega$ is exact, which shows that $\T\omega = 0$ from the unicity of the Hodge decomposition. \hfill $\square$

\newpage
\onecolumngrid

\bibliographystyle{QV_mnras}
\bibliography{paper_NC}

\begin{thebibliography}{}
\makeatletter
\relax
\def\mn@urlcharsother{\let\do\@makeother \do\$\do\&\do\#\do\^\do\_\do\%\do\~}
\def\mn@doi{\begingroup\mn@urlcharsother \@ifnextchar [ {\mn@doi@}
  {\mn@doi@[]}}
\def\mn@doi@[#1]#2{\def\@tempa{#1}\ifx\@tempa\@empty \href
  {http://dx.doi.org/#2} {doi:#2}\else \href {http://dx.doi.org/#2} {#1}\fi
  \endgroup}
\def\mn@eprint#1#2{\mn@eprint@#1:#2::\@nil}
\def\mn@eprint@arXiv#1{\href {http://arxiv.org/abs/#1} {{\tt arXiv:#1}}}
\def\mn@eprint@dblp#1{\href {http://dblp.uni-trier.de/rec/bibtex/#1.xml}
  {dblp:#1}}
\def\mn@eprint@#1:#2:#3:#4\@nil{\def\@tempa {#1}\def\@tempb {#2}\def\@tempc
  {#3}\ifx \@tempc \@empty \let \@tempc \@tempb \let \@tempb \@tempa \fi \ifx
  \@tempb \@empty \def\@tempb {arXiv}\fi \@ifundefined
  {mn@eprint@\@tempb}{\@tempb:\@tempc}{\expandafter \expandafter \csname
  mn@eprint@\@tempb\endcsname \expandafter{\@tempc}}}

\bibitem[\protect\citeauthoryear{{Barrow}}{{Barrow}}{2020}]{2020_Barrow}
{Barrow} J.~D.,  2020, { \it {Non-Euclidean Newtonian cosmology}}, \mn@doi
  [\textcolor{LinkJournal}{Classical and Quantum Gravity}]
  {10.1088/1361-6382/ab8437}, \href
  {https://ui.adsabs.harvard.edu/abs/2020CQGra..37l5007B}
  {\textcolor{LinkADS}{37}}, \href {http://arxiv.org/abs/2002.10155}
  {\textcolor{LinkArXiv}{125007}}

\bibitem[\protect\citeauthoryear{{Buchert}}{{Buchert}}{2008}]{2008_Buchert}
{Buchert} T.,  2008, { \it {Dark Energy from structure: a status report}},
  \mn@doi [\textcolor{LinkJournal}{General Relativity and Gravitation}]
  {10.1007/s10714-007-0554-8}, \href
  {https://ui.adsabs.harvard.edu/abs/2008GReGr..40..467B}
  {\textcolor{LinkADS}{40}}, \href {http://arxiv.org/abs/0707.2153}
  {\textcolor{LinkArXiv}{467}}

\bibitem[\protect\citeauthoryear{{Buchert}}{{Buchert}}{2018}]{2018_Buchert}
{Buchert} T.,  2018, { \it {On Backreaction in Newtonian cosmology}}, \mn@doi
  [\textcolor{LinkJournal}{\mnras}] {10.1093/mnrasl/slx160}, \href
  {https://ui.adsabs.harvard.edu/abs/2018MNRAS.473L..46B}
  {\textcolor{LinkADS}{473}}, \href {http://arxiv.org/abs/1704.00703}
  {\textcolor{LinkArXiv}{L46}}

\bibitem[\protect\citeauthoryear{{Buchert} \& {Ehlers}}{{Buchert} \&
  {Ehlers}}{1997}]{1997_Buchert_et_al}
{Buchert} T.,  {Ehlers} J.,  1997, { \it {Averaging inhomogeneous Newtonian
  cosmologies.}},
  \href{http://aa.springer.de/bibs/7320001/2300001/small.htm}{\textcolor{LinkJournal}{\aap}},
  \href {https://ui.adsabs.harvard.edu/abs/1997A&A...320....1B}
  {\textcolor{LinkADS}{320}}, \href {http://arxiv.org/abs/astro-ph/9510056}
  {\textcolor{LinkArXiv}{1}}

\bibitem[\protect\citeauthoryear{{Buchert} \& {M{\"a}dler}}{{Buchert} \&
  {M{\"a}dler}}{2019}]{2019_Buchert_et_al}
{Buchert} T.,  {M{\"a}dler} T.,  2019, { \it {Editorial note to: On the
  Newtonian limit of Einstein's theory of gravitation (by J{\"u}rgen Ehlers)}},
  \mn@doi [\textcolor{LinkJournal}{General Relativity and Gravitation}]
  {10.1007/s10714-019-2623-1}, \href
  {https://ui.adsabs.harvard.edu/abs/2019GReGr..51..162B}
  {\textcolor{LinkADS}{51}}, \href {http://arxiv.org/abs/1910.12106}
  {\textcolor{LinkArXiv}{162}}

\bibitem[\protect\citeauthoryear{Cartan}{Cartan}{1923}]{1923_Cartan}
Cartan E.,  1923, { \it {Sur les vari\'et\'es \`a connexion affine et la
  th\'eorie de la relativit\'e g\'en\'eralis\'ee (premi\`ere partie)}}, \mn@doi
  [\textcolor{LinkJournal}{Annales scientifiques de l'\'Ecole Normale
  Sup\'erieure}] {10.24033/asens.751}, 3e s{\'e}rie, 40, 325

\bibitem[\protect\citeauthoryear{Cartan}{Cartan}{1924}]{1924_Cartan}
Cartan E.,  1924, { \it Sur les vari\'et\'es \`a connexion affine, et la
  th\'eorie de la relativit\'e g\'en\'eralis\'ee (premi\`ere partie) (Suite)},
  \mn@doi [\textcolor{LinkJournal}{Annales scientifiques de l'\'Ecole Normale
  Sup\'erieure}] {10.24033/asens.753}, 3e s{\'e}rie, 41, 1

\bibitem[\protect\citeauthoryear{{Dautcourt}}{{Dautcourt}}{1990a}]{1990_Dautcourt_a}
{Dautcourt} G.,  1990a, { \it {On the Newtonian limit of general relativity}},
  \href{https://www.actaphys.uj.edu.pl/R/21/10/755}{\textcolor{LinkJournal}{Acta
  Phys. Pol. B}}, 21, 755

\bibitem[\protect\citeauthoryear{{Dautcourt}}{{Dautcourt}}{1990b}]{1990_Dautcourt_b}
{Dautcourt} G.,  1990b, { \it {Cosmological Coriolis fields in the
  Newton-Cartan theory}}, \mn@doi [\textcolor{LinkJournal}{General Relativity
  and Gravitation}] {10.1007/BF00764155}, \href
  {https://ui.adsabs.harvard.edu/abs/1990GReGr..22..765D}
  {\textcolor{LinkADS}{22}}, 765

\bibitem[\protect\citeauthoryear{{Dautcourt}}{{Dautcourt}}{1997}]{1997_Dautcourt}
{Dautcourt} G.,  1997, { \it {Post-Newtonian extension of the Newton - Cartan
  theory}}, \mn@doi [\textcolor{LinkJournal}{Classical and Quantum Gravity}]
  {10.1088/0264-9381/14/1A/009}, \href
  {https://ui.adsabs.harvard.edu/abs/1997CQGra..14A.109D}
  {\textcolor{LinkADS}{14}}, \href {http://arxiv.org/abs/gr-qc/9610036}
  {\textcolor{LinkArXiv}{A109}}

\bibitem[\protect\citeauthoryear{{Ehlers}}{{Ehlers}}{2019}]{2019_Ehlers}
{Ehlers} J.,  2019, { \it {Republication of: On the Newtonian limit of
  Einstein’s theory of gravitation}}, \mn@doi
  [\textcolor{LinkJournal}{General Relativity and Gravitation}]
  {10.1007/s10714-019-2624-0}, 51

\bibitem[\protect\citeauthoryear{{Ellis}}{{Ellis}}{1967}]{1967_Ellis}
{Ellis} G.~F.~R.,  1967, { \it {Dynamics of Pressure-Free Matter in General
  Relativity}}, \mn@doi [\textcolor{LinkJournal}{Journal of Mathematical
  Physics}] {10.1063/1.1705331}, \href
  {https://ui.adsabs.harvard.edu/abs/1967JMP.....8.1171E}
  {\textcolor{LinkADS}{8}}, 1171

\bibitem[\protect\citeauthoryear{{Kaiser}}{{Kaiser}}{2017}]{2017_Kaiser}
{Kaiser} N.,  2017, { \it {Why there is no Newtonian backreaction}}, \mn@doi
  [\textcolor{LinkJournal}{\mnras}] {10.1093/mnras/stx907}, \href
  {https://ui.adsabs.harvard.edu/abs/2017MNRAS.469..744K}
  {\textcolor{LinkADS}{469}}, \href {http://arxiv.org/abs/1703.08809}
  {\textcolor{LinkArXiv}{744}}

\bibitem[\protect\citeauthoryear{K{\"u}nzle}{K{\"u}nzle}{1972}]{1972_Kunzle}
K{\"u}nzle H.~P.,  1972, { \it Galilei and Lorentz structures on space-time :
  comparison of the corresponding geometry and physics},
  \href{http://www.numdam.org/item/AIHPA_1972__17_4_337_0}{\textcolor{LinkJournal}{Annales
  de l'I.H.P. Physique th{\'e}orique}}, 17, 337

\bibitem[\protect\citeauthoryear{{K{\"u}nzle}}{{K{\"u}nzle}}{1976}]{1976_Kunzle}
{K{\"u}nzle} H.~P.,  1976, { \it {Covariant Newtonian limit of Lorentz
  space-times}}, \mn@doi [\textcolor{LinkJournal}{General Relativity and
  Gravitation}] {10.1007/BF00766139}, \href
  {https://ui.adsabs.harvard.edu/abs/1976GReGr...7..445K}
  {\textcolor{LinkADS}{7}}, 445

\bibitem[\protect\citeauthoryear{{Misner}, {Thorne}  \& {Wheeler}}{{Misner}
  et~al.}{1973}]{1973_MTW}
{Misner} C.~W.,  {Thorne} K.~S.,   {Wheeler} J.~A.,  1973, { \it Gravitation},
  San Francisco: W.H. Freeman and Co.

\bibitem[\protect\citeauthoryear{{Roukema} \& {R{\'o}{\.z}a{\'n}ski}}{{Roukema}
  \& {R{\'o}{\.z}a{\'n}ski}}{2009}]{2009_Roukema_et_al}
{Roukema} B.~F.,  {R{\'o}{\.z}a{\'n}ski} P.~T.,  2009, { \it {The residual
  gravity acceleration effect in the Poincar{\'e} dodecahedral space}}, \mn@doi
  [\textcolor{LinkJournal}{\aap}] {10.1051/0004-6361/200911881}, \href
  {https://ui.adsabs.harvard.edu/abs/2009A&A...502...27R}
  {\textcolor{LinkADS}{502}}, \href {http://arxiv.org/abs/0902.3402}
  {\textcolor{LinkArXiv}{27}}

\bibitem[\protect\citeauthoryear{{Roy}}{{Roy}}{2014}]{2014_Roy}
{Roy} X.,  2014, { \it {On the 1+3 Formalism in General Relativity}}, arXiv
  e-prints, \href {https://ui.adsabs.harvard.edu/abs/2014arXiv1405.6319R}
  {\textcolor{LinkADS}{\href
  {https://ui.adsabs.harvard.edu/abs/2014arXiv1405.6319R}
  {\textcolor{LinkADS}{ADS link}}}}, \href {http://arxiv.org/abs/1405.6319}
  {\textcolor{LinkArXiv}{arXiv:1405.6319}}

\bibitem[\protect\citeauthoryear{{Ruede} \& {Straumann}}{{Ruede} \&
  {Straumann}}{1997}]{1997_Ruede_et_al}
{Ruede} C.,  {Straumann} N.,  1997, { \it {On Newton-Cartan cosmology.}},
  \href{https://www.e-periodica.ch/digbib/view?pid=hpa-001\%3A1997\%3A70\%3A\%3A3#324}{\textcolor{LinkJournal}{Helvetica
  Physica Acta}}, \href {https://ui.adsabs.harvard.edu/abs/1997AcHPh..70..318R}
  {\textcolor{LinkADS}{70}}, \href {http://arxiv.org/abs/gr-qc/9604054}
  {\textcolor{LinkArXiv}{318}}

\bibitem[\protect\citeauthoryear{{Straumann}}{{Straumann}}{2008}]{2008_Straumann}
{Straumann} N.,  2008, { \it {Proof of a decomposition theorem for symmetric
  tensors on spaces with constant curvature}}, \mn@doi
  [\textcolor{LinkJournal}{Annalen der Physik}] {10.1002/andp.200810312}, \href
  {https://ui.adsabs.harvard.edu/abs/2008AnP...520..609S}
  {\textcolor{LinkADS}{520}}, \href {http://arxiv.org/abs/0805.4500}
  {\textcolor{LinkArXiv}{609}}

\bibitem[\protect\citeauthoryear{{Tafel}}{{Tafel}}{2018}]{2018_Tafel}
{Tafel} J.,  2018, { \it {All transverse and TT tensors in flat spaces of any
  dimension}}, \mn@doi [\textcolor{LinkJournal}{General Relativity and
  Gravitation}] {10.1007/s10714-018-2355-7}, \href
  {https://ui.adsabs.harvard.edu/abs/2018GReGr..50...31T}
  {\textcolor{LinkADS}{50}}, \href {http://arxiv.org/abs/1710.10605}
  {\textcolor{LinkArXiv}{31}}

\bibitem[\protect\citeauthoryear{{Tichy} \& {Flanagan}}{{Tichy} \&
  {Flanagan}}{2011}]{2011_Tichy_et_al}
{Tichy} W.,  {Flanagan} {\'E}.~{\'E}.,  2011, { \it {Covariant formulation of
  the post-1-Newtonian approximation to general relativity}}, \mn@doi
  [\textcolor{LinkJournal}{\prd}] {10.1103/PhysRevD.84.044038}, \href
  {https://ui.adsabs.harvard.edu/abs/2011PhRvD..84d4038T}
  {\textcolor{LinkADS}{84}}, \href {http://arxiv.org/abs/1101.0588}
  {\textcolor{LinkArXiv}{044038}}

\bibitem[\protect\citeauthoryear{{Toupin}}{{Toupin}}{1957}]{1957_Toupin}
{Toupin} R.~A.,  1957, { \it {World invariant kinematics}}, \mn@doi
  [\textcolor{LinkJournal}{Archive for Rational Mechanics and Analysis}]
  {10.1007/BF00298004}, \href
  {https://ui.adsabs.harvard.edu/abs/1957ArRMA...1..181T}
  {\textcolor{LinkADS}{1}}, 181

\bibitem[\protect\citeauthoryear{Trautman}{Trautman}{1963}]{1963_trautman}
Trautman A.,  1963, { \it Sur la théorie newtonienne de la gravitation},
  \href{https://gallica.bnf.fr/ark:/12148/bpt6k4007z/f639.image}{\textcolor{LinkJournal}{C.
  R. Acad. Sci. (Paris)}}, 257

\bibitem[\protect\citeauthoryear{{Vigneron}}{{Vigneron}}{2020}]{2020_Vigneron}
{Vigneron} Q.,  2020, { \it {1 +3 formulation of Newton's equations}}, \mn@doi
  [\textcolor{LinkJournal}{\prd}] {10.1103/PhysRevD.102.124005}, \href
  {https://ui.adsabs.harvard.edu/abs/2020PhRvD.102l4005V}
  {\textcolor{LinkADS}{102}}, \href {http://arxiv.org/abs/2010.10247}
  {\textcolor{LinkArXiv}{124005}}

\makeatother
\end{thebibliography}

\vspace{1cm}

{When available this bibliography style features three different links associated with three different colors: links to the journal or editor website are in \textcolor{LinkJournal}{red}, links to the ADS website are in \textcolor{LinkADS}{blue} and links to the arXiv website are in \textcolor{LinkArXiv}{green}.}

\end{document}